\documentclass[aps,pre,twocolumn,groupaddress]{revtex4-1}
\usepackage{color}
\usepackage{graphicx}

\begin{document}
\title{Arcsine Law and Multistable Brownian Dynamics in a Tilted Periodic Potential}
%
\author{J. Spiechowicz and J. {\L}uczka}
\affiliation{Institute of Physics, University of Silesia, 41-500 Chorz{\'o}w, Poland}
\begin{abstract}
Multistability is one of the most important phenomena in dynamical systems, e.g. bistability enables the implementation of logic gates and therefore computation. Recently multistability has attracted a greatly renewed interest related to memristors and graphene structures, to name only a few. We investigate {\it tristability} in velocity dynamics of a Brownian particle subjected to a tilted periodic potential. It is demonstrated that the origin of this effect is attributed 
the arcsine law for the velocity dynamics at the zero temperature limit. We analyze the impact of thermal fluctuations and construct the phase diagram for the stability of the velocity dynamics. 
It 
suggests an efficient strategy to control the multistability by changing solely the force acting on the particle or temperature of the system.  Our findings for the paradigmatic model of nonequilibrium statistical physics apply to, \emph{inter alia}, Brownian motors, Josephson junctions, cold atoms dwelling in optical lattices and colloidal systems.
\end{abstract}
\maketitle
\section{Introduction}
The Brownian motion of a particle dwelling in a tilted periodic potential has served as an  archetypal model of transport phenomena in nonequilibrium statistical physics that for decades has been applied to both classical and quantum systems. Examples are the dynamics of the phase difference across Josephson junctions \cite{junction}, rotating dipoles in external fields \cite{coffey}, superionic conductors \cite{fulde1975}, charge density waves \cite{gruner1981} and cold atoms dwelling in optical lattices \cite{denisov2014,kindermann2017,dechant2019}, to mention only a few. Such a prominent model can be formulated in terms of the following dimensionless Langevin equation
\begin{equation}
	\label{dimless-model}
	\ddot{x} + \gamma \dot{x} = -\mathcal{U}'(x) + \sqrt{2\gamma \theta}\,\xi(t),
\end{equation}
where $\gamma$ is the Stokes friction coefficient, 
$\mathcal{U}(x) = -\sin{x} - fx$ is the total potential, 
$f$ the constant bias and $\theta \propto k_B T$ is the dimensionless temperature of the system. The coupling of the particle with thermostat is modeled by the $\delta$-correlated Gaussian white noise $\xi(t)$ of vanishing mean, namely $\langle \xi(t) \rangle = 0$ and \mbox{$\langle \xi(t)\xi(s) \rangle = \delta(t-s)$}. The starting dimensional equation is presented in Appendix A, where the corresponding scaling in defined as well.

\begin{figure}[b]
	\centering
	\includegraphics[width=0.9\linewidth]{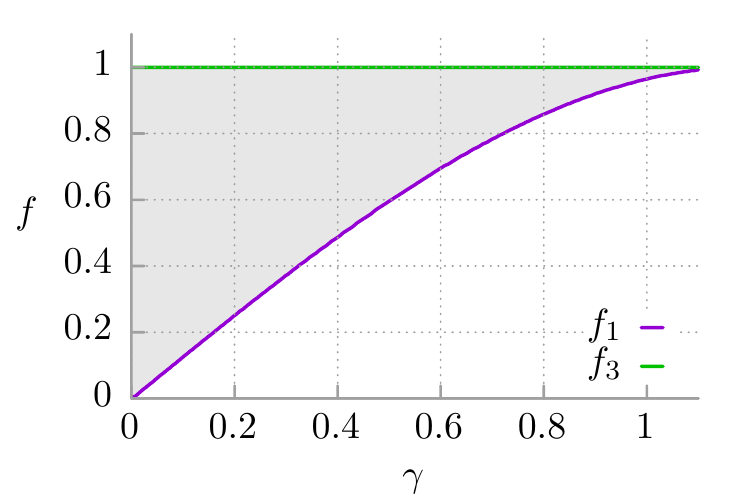}
	\caption{Phase diagram for occurrence of the velocity bistability phenomenon in the deterministic counterpart of the system (1) presented in the dimensionless parameter plane $(\gamma,f)$. The critical forces $f_1(\gamma)$ as well as $f_3$ are indicated by the corresponding lines. Velocity bistability is observed only if the friction and force values lay in the marked gray area.}
	\label{fig1}
\end{figure}

This nonlinear stochastic system enjoys seemingly never-ending interest as its different aspects have been studied already for several decades \cite{festa, risken, lindner2001, reimann2001a, reimann2001b, constantini1999, lindenberg2005, kramer2013, marchenko2014, lindner2016, zhang2017, cheng2018, goychuk2019, spiechowicz2020pre, goychuk2020, li2020, spiechowicz2020pre2}. For example it may exhibit unusual phenomena like the giant diffusion \cite{lindner2001,reimann2001a,reimann2001b,lindner2016} or non-monotonic temperature dependence of the diffusion coefficient \cite{spiechowicz2020pre}. These effects have been associated with  bistability of the velocity dynamics, the phenomenon  well known since the seminal work of Risken et al \cite{vollmer1983} who first found that at low friction and appropriate bias values the velocity $v(t)=\dot x(t)$ can be stable in a locked state (the particle is trapped in a potential minimum) and also in a running one (the motion is unbounded in space). To observe this phenomenon in the deterministic system the constant force $f$ should be in range $f_1(\gamma)< f < f_3 = 1$, where the friction-dependent boundary $f_1$ is the minimal value $f$ for which the running state starts to appear while for $f > f_3 = 1$ they occur exclusively. In Fig. \ref{fig1} we present the phase diagram for the emergence of the velocity bistability \cite{risken}. This effect is observed only if the friction and force values lay in the marked gray region. 

In this paper, we focus on tristability of the Brownian velocity dynamics.  Despite many years of intense and beneficial studies, this distinctive feature in the considered system has been addressed only recently \cite{spiechowicz2020pre,ivan}. This effect has also been reported in similar setups with fractional hydrodynamic memory \cite{goychuk2019}, for particles subjected to nonlinear friction and driven by Ornstein-Uhleneck L{\`e}vy noise \cite{li2020} and also in systems perturbed by the harmonic L{\`e}vy process \cite{li2021}. Other systems exhibiting tristability are currently under intense investigations. Examples are Kerr cavities driven by optical waves \cite{murdoch}, viscoelastic flows through microscale porous arrays \cite{shen} and graphene structures \cite{zhao}. 

The structure of the paper is as follows. In Sec. II the problem of velocity tristability in noisy system is introduced. Next, to explain the origin of this effect we first study the dynamics of the deterministic system. Later we discuss the impact of thermal fluctuations on  stability of the velocity dynamics. In. Sec. III we present the phase diagram for different temperature of the system. Finally, Sec. IV provides discussion of the results and a summary. In the Appendixes we present the scaling of the Langevin equation describing the system, discuss on its ergodicity and recall the arcsine distribution. 
\begin{figure}[t]
	\centering
	\includegraphics[width=0.9\linewidth]{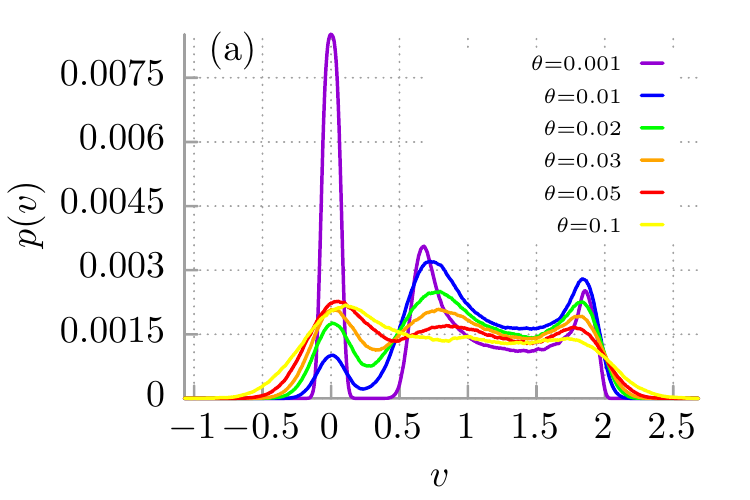}\\
	\includegraphics[width=0.9\linewidth]{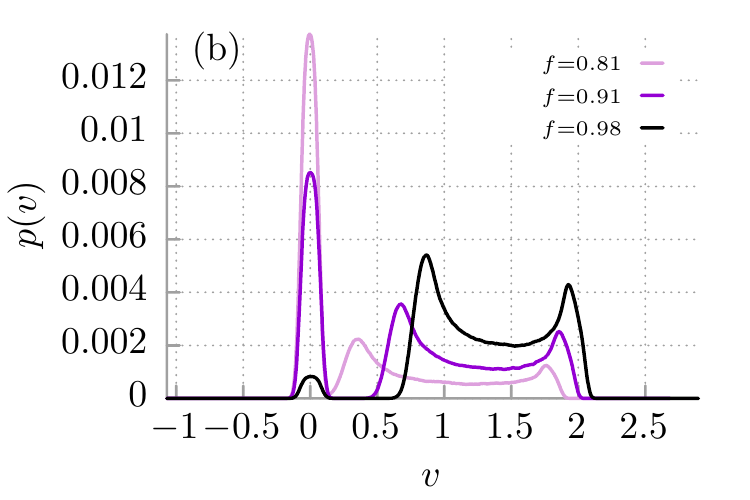}
	\caption{The probability distribution $p(v)$ for the instantaneous long time velocity $v$ of the Brownian particle is depicted in panel (a) for selected values of the dimensionless temperature $\theta$. Parameters are $\gamma = 0.66$ and $f = 0.91$. In panel (b) the same characteristic is shown but for fixed temperature $\theta = 0.001$ and varied bias $f$ values.}
	\label{fig2}
\end{figure}

\section{Tristability of velocity dynamics}
At first glance the dynamics described by Eq. (\ref{dimless-model}) may look simple, but the Fokker-Planck equation for the particle probability distribution $P(x,\dot{x},t)$ corresponding to it is a second order partial differential equation in three variables $(x, \dot{x}, t)$ whose exact solutions up to now are not known. Moreover, the parameter space of the model given by Eq. (\ref{dimless-model}) is multidimensional $\{\gamma, f, \theta\}$. Therefore we used extensive numerical simulations of the Langevin dynamics to analyze the stability of velocity dynamics.

All numerical calculations have been done using a Compute Unified Device Architecture (CUDA) environment implemented on a modern desktop Graphics Processing Unit (GPU). This method allowed us to speedup necessary calculations by a factor of the order $10^3$ as compared to the present Central Processing Unit (CPU) schemes \cite{spiechowicz2015cpc}. In Appendix B we show that in the deterministic limit $\theta = 0$ the ergodicity of the studied system is strongly broken, i.e. its phase space is decomposed onto two non-intersecting invariant corresponding to the locked and running states. Consequently, to get rid of the dependence of the obtained results on the initial position $x(0) = x_0$ and velocity $v(0) = v_0$ it is necessary to perform additional average over them. They were uniformly distributed on the intervals $[0,2\pi]$ and $[-2,2]$, respectively. We note that this procedure embodies many of the experimental situations where the initial conditions are not known \emph{a priori}. For any non-zero temperature $\theta > 0$ the ergodicity of the system is restored \cite{pavliotis,cheng} and the results are no longer dependent on the initial conditions. Unless stated otherwise the statistical quantities characterizing the transport properties of the system were calculated over the ensemble of $2^{19} = 524288$ trajectories, each starting with different initial conditions $\{x_0, v_0\}$  according to the above distributions. All so obtained results have been tested for convergence with respect to (i) a time step of numerical integration; (ii) a size of the statistical ensemble; (iii) a volume of initial conditions. 


In Fig. \ref{fig2}  we exemplify the velocity tristability phenomenon.  The probability distribution $p(v)$ for the instantaneous long time velocity $v$ of the particle is trimodal, i.e. it exhibits three maxima: one corresponding to the locked state $v=0$ and the remaining two related to the running state $v \ne 0$.  The latter fact means that the values of instantaneous velocity $v$ close to these extrema occur significantly more often than others and in this sense are more stable \cite{grebogi}.
\begin{figure}[t]
	\centering
	\includegraphics[width=0.9\linewidth]{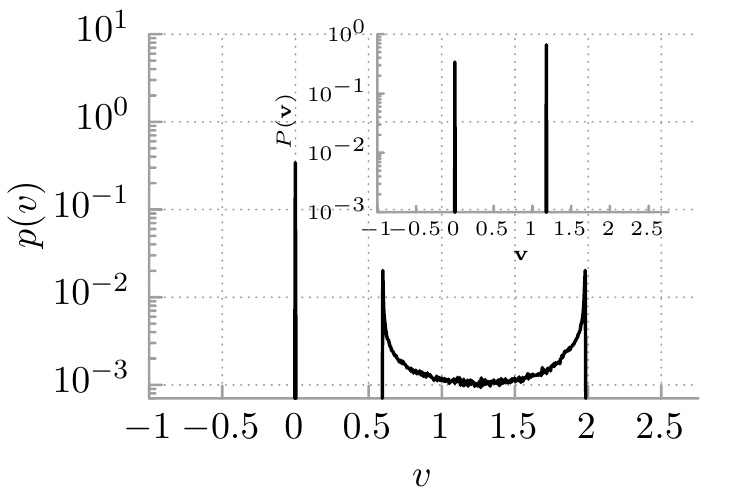}
	\caption{The probability distribution $p(v)$ for the instantaneous long time velocity $v$ of the deterministic system $\theta = 0$. The maxima of $p(v)$ are at $v_1=0$, $v_m = 0.59$ and $v_M=1.98$. In the inset we display the corresponding distribution $P(\mathbf{v})$ for the time averaged velocity $\mathbf{v}$. The parameters are $\gamma = 0.66$ and $f = 0.91$.}
	\label{fig3}
\end{figure}

\subsection{Deterministic system, $\theta = 0$}
The nature of tristability in the velocity Brownian dynamics has not been explained. It often happens that the mechanism behind a physical effect is hidden in the simplified deterministic behavior of a system \cite{slapik2018,slapik2019,spiechowicz2019njp}. We follow this route and start our analysis with the deterministic counterpart of the system, i.e. when  $\theta = 0$ in Eq. (\ref{dimless-model}). In \mbox{Fig. \ref{fig3}} we depict the probability distribution $p(v)$ for instantaneous velocity $v$ at the fixed point of time $t = t_i \gg 1$ (in simulations we fixed $t_i=10^5$). It corresponds to the parameters regime presented in panel (a) of Fig. \ref{fig2}.  Three pronounced peaks are visible there. The first $v_1 = 0$ describes the locked state whereas the remaining two $v_m \approx 0.59$ and $v_M \approx 1.98$ in $U$-shaped part of $p(v)$ correspond to the running state. Therefore already in the deterministic limit $\theta = 0$ the actual velocity dynamics is not bistable but rather tristable. Note that these three peaks are directly related to the position of three maxima of the distribution $p(v)$ for non-zero temperature $\theta > 0$ shown in Fig. \ref{fig2} (a). Now it is clear that the tristability of velocity observed for the noisy system must be rooted in the deterministic dynamics.
\begin{figure}[t]
	\centering
	\includegraphics[width=0.9\linewidth]{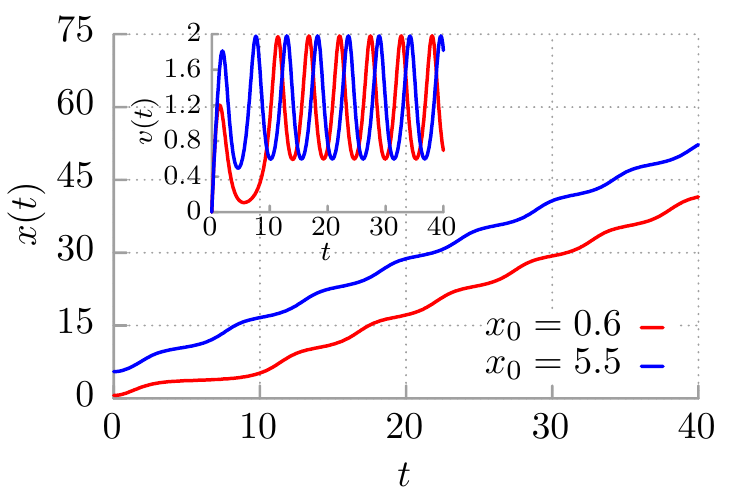}
	\caption{The exemplary deterministic trajectories of the particle coordinate $x(t)$ and velocity $\dot{x}(t) = v(t)$ (in the inset) vs time  for two different initial conditions $x_0$ corresponding to the running state. Other parameters are $\gamma = 0.66$, $f = 0.91$, $\theta = 0$ and $v_0 = 0$.}
	\label{fig4}
\end{figure} 

In the inset of Fig. \ref{fig3} we illustrate the velocity bistability in terms of the probability distribution $P(\mathbf{v})$ of the time-averaged velocity $\mathbf{v} = \lim_{t \to \infty} (1/t) \int_0^t \dot x(s) \, ds$. After the temporal averaging the continuum of instantaneous velocities $v$ in the $U$-shaped part of $p(v)$ collapses onto the single attractor for the time-averaged velocity $\mathbf{v} \approx 1.18$. We stress that for the asymptotic long time regime the probability measure $p(v)$ is time-invariant and therefore the choice of time instant $t_i \gg 1$  is arbitrary. For  $\gamma =0.66$ and $f=0.91$  the support of the $U$-shaped part of $p(v)$ corresponding to the running state is the interval $[v_m, v_M]=[0.59,1.98]$. Moreover, the maxima of $p(v)$ are located at the boundary of this window. If for the fixed $\gamma$ the force $f$ increases then both $v_m$ and $v_M$ grow, but $|v_M - v_m|$ decreases, c.f. Fig. \ref{fig2} (b). In turn, if for the fixed force $f$ the friction $\gamma$ grows then both $v_m$ and $v_M$ decrease and at the same time $|v_M - v_m|$ increases. 
\begin{figure}[t]
	\centering
	\includegraphics[width=0.9\linewidth]{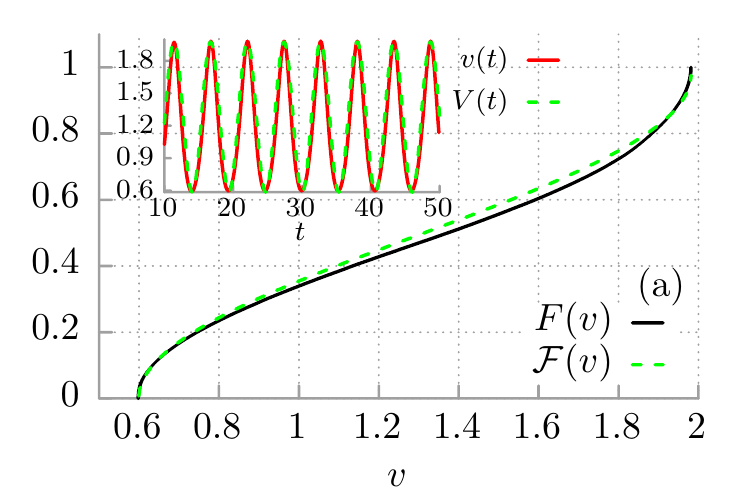}\\
	\includegraphics[width=0.9\linewidth]{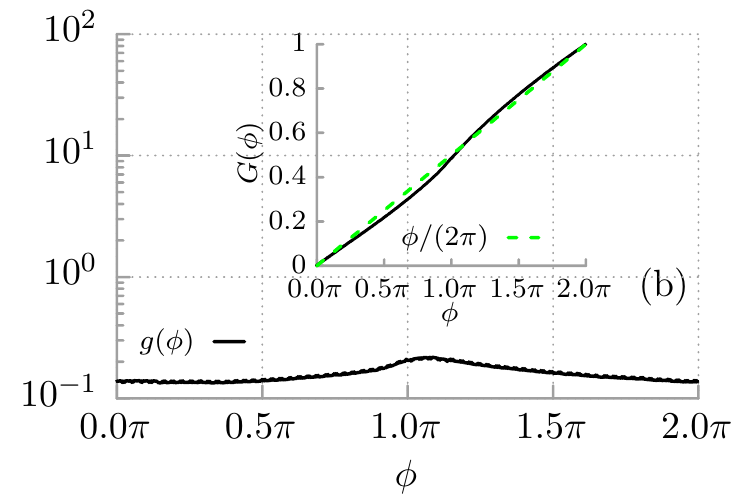}
	\caption{Panel (a): The cumulative distribution function $F(v)$ for the instantaneous long time velocity $v$ corresponding to the $U$-shaped part of the density $p(v)$ is compared with the arcsine distribution ${\mathcal F}(v) = (2/\pi) \arcsin [\sqrt{(v-v_m)/(v_M - v_m)}]$. 
	Inset: The solution $v(t)$ for $x_0 = 0.6$ and $v_0 = 0$ is compared  with the function $V(t) = A\sin{(\omega t + \phi)} + c$, where $A = 0.695$, $\omega = 1.18$, $\phi = -5.6$ and $c = 1.28$. Other parameters are $\gamma = 0.66$, $f = 0.91$ and $\theta = 0$.
	Panel (b): The probability density $g(\phi)$ for the phase difference $\phi$ between the trajectories $v(t)$ in the long time regime. 
	Inset: The cumulative distribution function $G(\phi)$ corresponding to the density $g(\phi)=dG(\phi)/d\phi$ is compared with the cumulative distribution function of the uniformly distributed random variable (the green curve). Parameters are $\gamma = 0.66$, $f = 0.91$ and $\theta = 0$.}
	\label{fig5}
\end{figure}

In order to explain the origin of the $U$-shaped part of $p(v)$, in Fig. \ref{fig4} we display two exemplary deterministic trajectories for the position $x(t)$ and velocity $v(t)$ corresponding to the running state. For any initial conditions $\{x_0,v_0\}$ corresponding to the running state (see Fig. \ref{fig1A} in Appendix B) the slope of the curves $x(t)$ is the same as the time averaged velocity $\mathbf{v}\approx 1.18$. However, the key factor for multistability are velocity trajectories of the running state: for long time $v(t)$ are time-periodic functions of the same period and amplitude but different phases which depend on $\{x_0, v_0\}$. The minimal and maximal values of velocity are $v_m = 0.59$ and $v_M = 1.98$, respectively, i.e. where the $U$-shaped part of $p(v)$ attains its maxima. We have found that the $U$-shaped part of the probability density $p(v)$ corresponding to the running state can be very well approximated by the arcsine distribution. In Fig. \ref{fig5} (a) we compare the cumulative distribution function $F(v)$ for the instantaneous velocity defined as 
\mbox{$F(v) = \int_{v_m}^v p(u) du$}
with the exact arcsine distribution given by the formula \mbox{${\mathcal F}(v) = (2/\pi) \arcsin [\sqrt{(v-v_m)/(v_M - v_m)}]$}. The agreement between these two curves is indeed quite remarkable. 
\begin{figure}[t]
	\centering
	\includegraphics[width=0.9\linewidth]{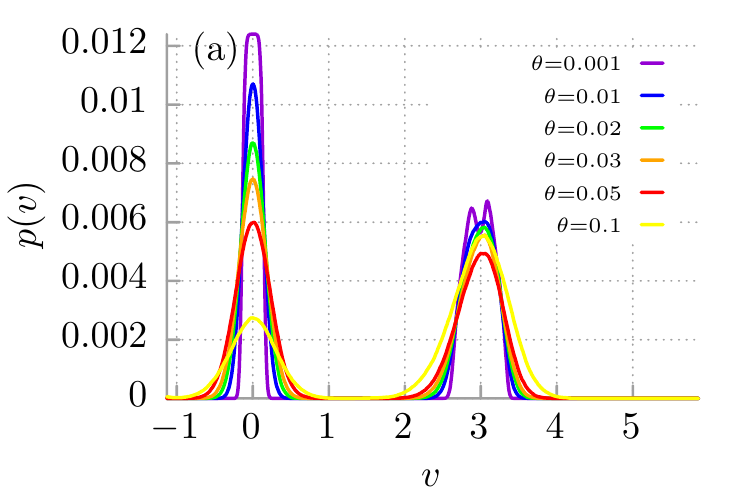}\\
	\includegraphics[width=0.9\linewidth]{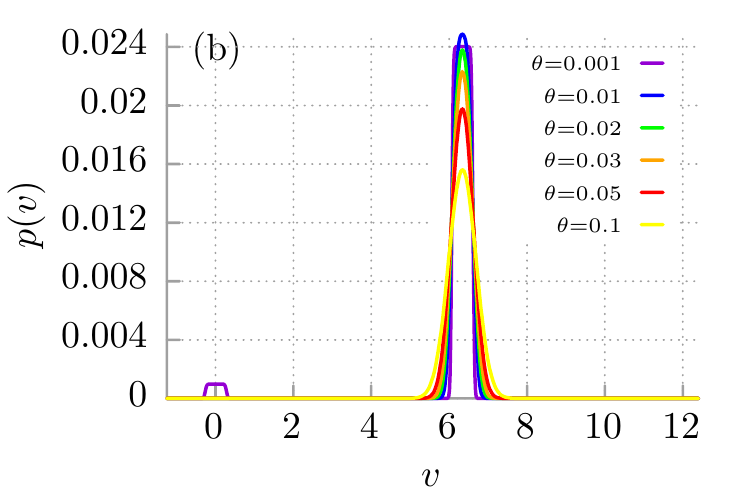}\\
	\includegraphics[width=0.9\linewidth]{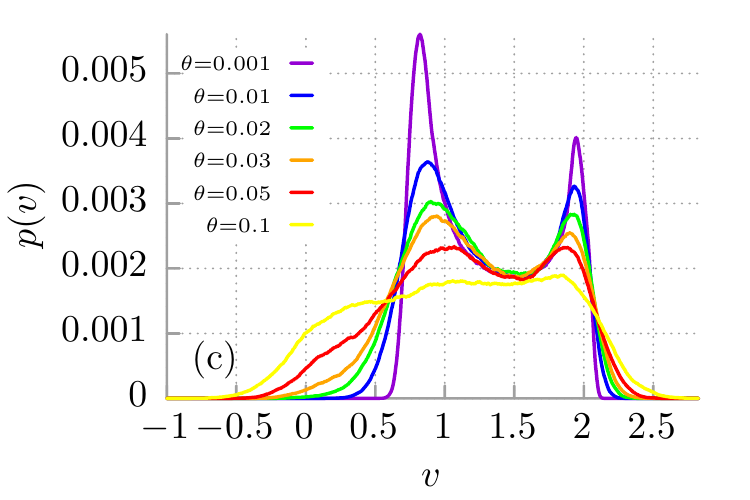}
	\caption{Influence of temperature on the probability distribution $p(v)$ of the instantaneous long time velocity of the Brownian particle depicted for different parameter $(\gamma, f)$ regimes. Panel (a): velocity bistability (coexistence of the locked and running states) for $\gamma = 0.15$ and $f = 0.45$ ; (b): velocity monostability (the running state) for $\gamma = 0.15$ and $f = 0.95$; (c): velocity bistability (induced by the running solutions) for $\gamma = 0.7$ and $f = 0.99$. Velocity multistability (coexistence of one locked and two running peaks) is exemplified for $\gamma = 0.66$ and $f = 0.91$ in Fig. \ref{fig2} (a).} 
	\label{fig3A}
\end{figure}

The origin of the arcsine law can be understood on the following basis. In the deterministic limit the asymptotic long time velocity $v(t)$ is a time-periodic function of period $\mathcal{T}$,  which has one maximum and one minimum per the period. The value of $\mathcal{T}$ depends only on $\gamma$ and $f$ and not on the initial coordinate and velocity of the particle. The analytical form of $v(t)$ is not know. However, in the inset of Fig. \ref{fig5} (a) we compare  $v(t)$ with the simple harmonic function 
\begin{equation}
	\label{approx}
	V(t) = A\sin{(\omega t + \phi)} + c,
\end{equation}
where $c = const$, $A = (v_M - v_m)/2 = 0.695$, $\omega = 2\pi/\mathcal{T} = L/\mathcal{T} = \mathbf{v} = 1.18$, i.e. the frequency $\omega$ is equal to the time averaged velocity $\mathbf{v}$ corresponding to the running state ($L = 2\pi$ stands for the spatial period of the potential). For long time and a fixed set $(\gamma, f)$, the parameters $(A, \omega, c)$ are the same for all initial conditions $\{x_0, v_0\}$ which correspond to the running state,  c.f. Fig. \ref{fig4}.  In Appendix C we show that if the phase shift $\phi$ in the function $V(t)$ is uniformly distributed on the interval $[0, 2\pi]$ then for an arbitrary but fixed time $t=t_i$ the random variable $\eta = V(t_i)$ follows exactly the arcsine law ${\mathcal F}(v)$ \cite{lee}. The compatibility of the probability distribution $F(v)$ for the particle and the arcsine distribution $\mathcal{F}(v)$ is not mathematically exact because of two factors: (i) certainly the function given by Eq. (\ref{approx}) does not obey Eq. (\ref{dimless-model}) and (ii) as it is shown in Fig. \ref{fig5} (b) the probability distribution of the phase shift is not strictly uniform. However, the "arcsine law" observed for the part of $p(v)$ corresponding to the running state is surprisingly good and explains its $U$-shape which in turn lies at the heart of the multistability of velocity dynamics in a tilted periodic potential.

Historically, the arcsine law was formulated by L{\`e}vy in 1940 for the distribution of the fraction of time that a trajectory of Brownian motion stays above zero and the last time when it visits the origin \cite{levy}.
This law is a cornerstone of extreme-value statistics and has been applied to describe \emph{inter alia} conductance in disordered materials \cite{beenakker1997}, mean magnetization in spin systems \cite{godreche2001}, currents in stochastic thermodynamics \cite{barato2018}, fractional \cite{sadhu2018} and aging \cite{akimoto2020}  Brownian motion, balistic L{\`e}vy random walks \cite{barkai}, to name only a few. In clear contrast to its common formulation, here the arcsine-like  law portrays the velocity distribution rather than the fraction of time that a trajectory of Brownian motion spends in some regions. This fact serves as a seed for multistability of the velocity dynamics.

\subsection{Influence of temperature}
In Fig. \ref{fig2} (a) and Fig. \ref{fig3A} we present the influence of temperature on the probability distribution $p(v)$ for the instantaneous long time velocity of the Brownian particle for a selected set of parameters $(\gamma, f)$  illustrating the complexity of different stability regimes. In panel (a) of Fig. \ref{fig3A},  for $\gamma = 0.15$ and $f = 0.45$,  we depict the case when the velocity multistability occurs. 
In this regime the mean velocity of the deterministic running state is $\langle v \rangle = 2.98$ while $v_m = 2.75$ and $v_M = 3.2$. 
We note that for low temperature $\theta = 0.001$ in fact the particle velocity is tristable. For $\theta = 0.01$ fine details of the deterministic dynamics are washed out and bistability is observed. Further growth of temperature causes the disappearing of the locked state so that for sufficiently high temperature there is only one running state (not depicted). In panel (b) of this figure  we exemplify the velocity monostability effect which occurs for the parameter pair $\gamma = 0.15$ and $f = 0.95$. 
In this regime the mean velocity of the deterministic running state is $\langle v \rangle = 6.34$ while $v_m =  6.175$ and $v_M = 6.49$. 
Here in the low temperature regime $\theta = 0.001$ the velocity dynamics displays the bistability phenomenon but as it is illustrated for $\theta > 0.01$ only the running state survives. When temperature increases the maximum of $p(v)$ is lowered. All fine details of the deterministic dynamics have disappeared apart from the fact that the running state still occurs with relatively high probability.  In the regime presented in panel (c) only the running state is detected and $p(v)$ exhibits two well separated peaks which are the residual arcsine-like law. The corresponding parameter regime reads $\gamma = 0.7$ and $f = 0.99$. 
In this regime the mean velocity of the deterministic running state is $\langle v \rangle = 1.24$ while $v_m = 0.67$ and $v_M = 2$. 
If thermal noise intensity grows the corresponding maxima of  $p(v)$ are gradually flattened until temperature of the order $\theta = 0.1$ when they almost disappear so that the velocity dynamics is monostable. Finally, in panel (a) of Fig. \ref{fig2} we depict the case of the multistable dynamics which occurs for $\gamma = 0.66$ and $f = 0.91$. Here the low temperature regime $\theta = 0.001$ resembles the deterministic structure of states as there are three peaks. The first corresponds to the locked solution whereas the remaining two indicate the running state with the residual arcsine law. The main difference is that the influence of temperature on the stability of the locked solution is non-monotonic. When thermal noise intensity grows first the locked state becomes less stable as the corresponding maximum is rapidly decreased. However, in the panel we illustrate that further increase of temperature enhances the stability of the locked solution. The impact of thermal noise intensity on the running state is the same as in previous cases, namely, when temperature grows two maxima are flattened so that it becomes less pronounced. In the high temperature limit the probability distribution $p(v)$ presents only the running state (not depicted).

We summarize this part with the conclusion that in the parameter region $f_1(\gamma) < f < f_3 = 1$, where in the deterministic counterpart of the system the effect of velocity multistability occurs, for the low to moderate temperature there are four different regimes of the velocity dynamics: (i) bistability (coexistence of the locked and running states); (ii) monostability (a single running state); (iii) bistability (induced by two peaks corresponding to the running state); (iv) multistability (coexistence of one locked and two arcsine-law peaks).
\begin{figure*}[t]
	\centering
	\includegraphics[width=0.32\linewidth]{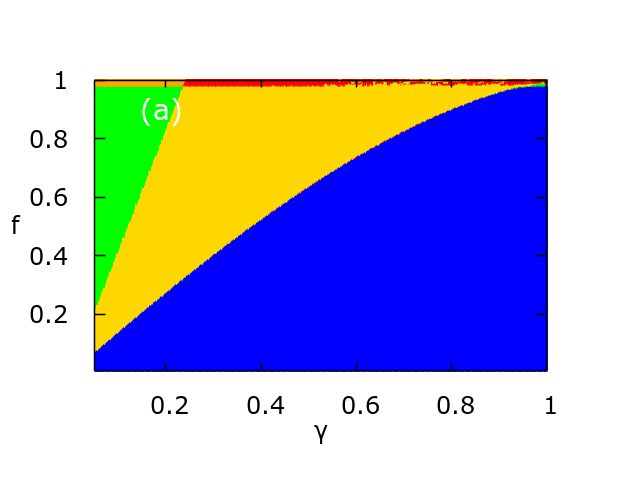}
	\includegraphics[width=0.32\linewidth]{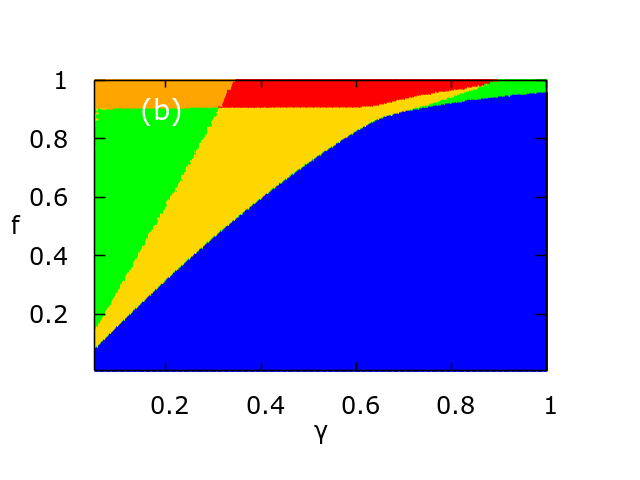}
	\includegraphics[width=0.32\linewidth]{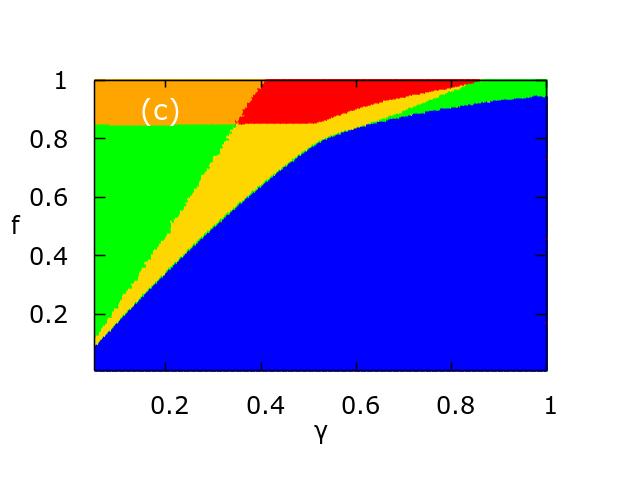}\\
	\includegraphics[width=0.32\linewidth]{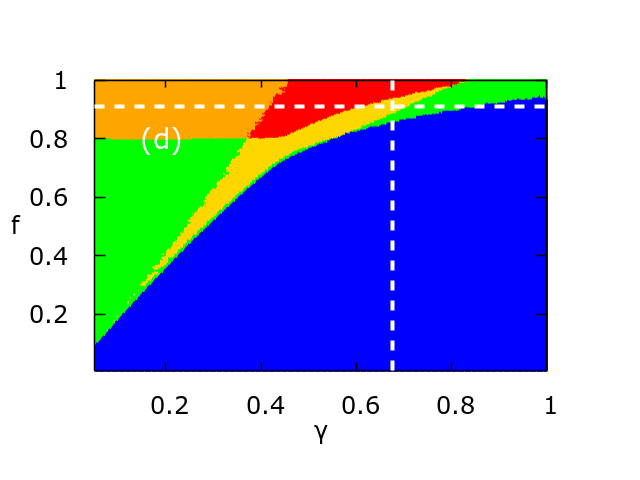}
	\includegraphics[width=0.32\linewidth]{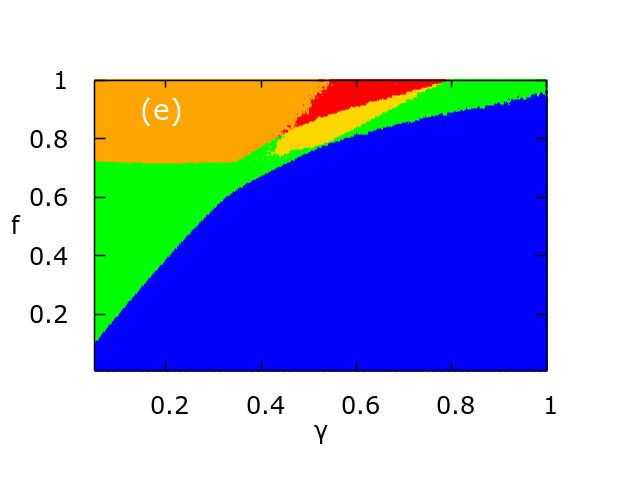}
	\includegraphics[width=0.32\linewidth]{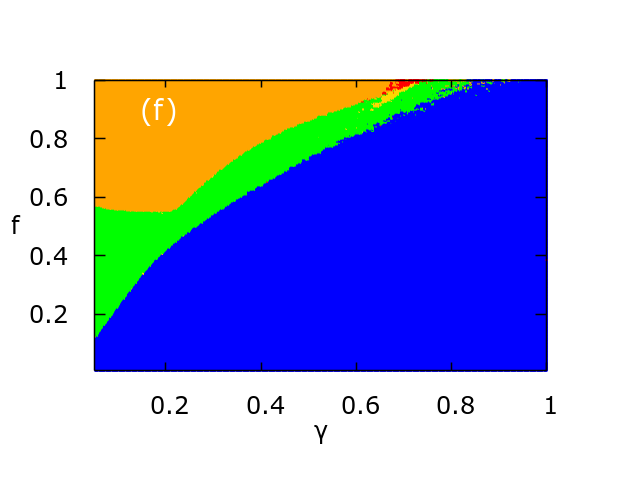}
	\caption{Phase diagram for the long time velocity dynamics of the Brownian particle dwelling in the tilted periodic potential shown in the parameter plane $(\gamma,f)$ for selected values of dimensionless temperature $\theta$. Panel (a): $\theta = 0.001$; (b): $\theta = 0.01$; (c): $\theta = 0.02$; (d): $\theta = 0.03$; (e): $\theta = 0.05$; (f): $\theta = 0.1$. Different regimes of the velocity states are marked with the corresponding color. Blue - the monostability (the single locked state). Orange - the monostability (the single running state). Green - the bistability (coexistence of the locked and running states). Red - the bistability (concurrence of two running peaks). Yellow - the multistability (coexistence of one locked and two running peaks).}
	\label{fig4A}
\end{figure*}

\section{Phase diagram}
Our innovative computational method of  integration of the Langevin equation (\ref{dimless-model}) allowed us to explore the parameter space with the unprecedented resolution in order to construct the phase diagram for the stability of the velocity dynamics. We performed scans of the following area in the parameter space $\gamma \times f \in [0,1] \times [0,1]$ for a given temperature $\theta$ at a resolution 400 points per dimension. For each pair $(\gamma, f)$ we calculated the probability distribution $p(v)$ for the instantaneous long time velocity and determined the regime of stability of the solutions, c.f. Fig. \ref{fig3A}. The result is presented in \mbox{Fig. \ref{fig4A}} where we show the phase diagrams for selected values of temperature $\theta$. Different regimes of the velocity dynamics are marked there with the corresponding color, namely, blue indicates the monostability (the single locked state); orange - the monostability (the single running state); green - the bistability (coexistence of the locked and running states); red - the bistability (two peaks corresponding to the running state); yellow - the multistability (coexistence of one locked and two arcsine-law peaks). It generalizes the phase diagram reported by Risken in his well-known book \cite{risken} by extending it to finite temperature regimes and including the effect of multistability.

A careful inspection e.g. of the Fig. \ref{fig4A} (d) reveals several regularities.  Firstly, close to the critical force $f_3 = 1$ there are only either one or two running states that translate to the mono or bistable regimes. Secondly, if $\gamma$ is sufficiently small then practically regardless of the bias $f$ the locked state coexists with the running state implying the bistability of the velocity dynamics. The same observation holds true when both $\gamma$ and $f$ are very close to unity. Thirdly, if $f < f_1(\gamma)$ then, as expected, only the locked state emerges. In between these areas there is a region where the trace of the deterministic dynamics in the form of the multistability is visible. The latter is particularly pronounced at low temperature. When thermal noise intensity grows this region is rapidly shrunk. As it is presented already for temperature $\theta = 0.1$ it ceases to exist. The same remark is valid also for the region denoted with red color that indicates the bistable dynamics with two concurrent running states. The latter effect occurs for moderate values of the friction coefficient $\gamma$ and the biasing force $f \approx 1$.

The phase diagrams presented in Fig. \ref{fig4A} suggests efficient strategies for controlling the multistability of the velocity dynamics. As it is demonstrated in Fig. \ref{fig2} (a) and Fig. \ref{fig3A} this goal can be achieved by altering temperature of the system. Such approach opens the opportunity to control the multistability of particles that carry no charge/dipole or can hardly be manipulated by means of an external field or force. For moderate temperature $\theta = 0.03$ one is able to cover all multistability regimes by fixing the force $f$ close to the critical bias $f = 1$ and changing the friction $\gamma$. We exemplify this case in Fig. \ref{fig4A} (d) by the white dashed horizontal line corresponding to $f = 0.91$. 
The most convenient way of manipulating the velocity stability regimes is by changing the constant force $f$ applied to the particle. It is illustrated in Fig. \ref{fig4A} (d) for $\gamma = 0.66$ indicated by the white dashed vertical line. For $f < 0.85$ there exist only single locked state (blue). In the interval $f = [0.85,1]$ first the bistability occurs (green) followed by the multistability (yellow) and then again the bistability but for the running state (red). 

Such an approach can be directly realized e.g. in a biased Josephson junction operating in a semi-classical regime described by the Stewart-McCumber model \cite{junction} in which the Josephson phase $\varphi(t)$ translates to the particle coordinate $x(t)$. Consequently, the voltage drop $\mathsf{V}(t) \propto \dot{\varphi}(t)$ across the device is equivalent to the velocity $v(t)$ whereas the constant current $\mathsf{I}$ applied to the junction translates to the bias acting on the particle $f$. Both the manufacturing of the Josephson junctions and generating of the DC are well-developed technologies nowadays and therefore our results are attainable experimentally and ready for corroboration.

\section{Discussion and summary}
In summary, we explained the underlying physical mechanism standing behind the emergence of multistability of random velocity dynamics of the Brownian particle in a tilted periodic potential. The origin of this effect lies in the time periodic dependence of the velocity $v(t)$ with almost uniformly distributed phase shift induced by different initial conditions for the position and velocity. The arcsine law for the particle velocity in the running state is not mathematically exact, however, we show that it is a very good approximation. With the locked state it generates the trimodal shape of the velocity distribution and in consequence the tristability of the Brownian particle velocity emerges. 

Moreover, we constructed the phase diagram for the stability of the velocity dynamics which generalizes the corresponding result \cite{risken} in a twofold way: it demonstrates the multistability of the velocity dynamics already in the deterministic system and secondly, describes the occurrence of the latter effect for non-zero temperature regimes. This phenomenon is restricted to the island on the $(\gamma, f)$-surface which is immersed in the region of bistability. As we demonstrated it can be continuously deformed by changing temperature of the system. Last but not least, it suggests the efficient strategy to control the multistability by changing solely e.g. the external bias force acting on the particle.

We expect that the presented mechanism can readily be experimentally verified in plentiful of different contexts and our findings can inspire a vibrant follow up work due to the universality and simplicity of the considered paradigmatic model of nonequilibrium statistical physics.

\section*{Acknowledgment}
This work has been supported by the Grant NCN No. 2017/26/D/ST2/00543 (J. S.)

\begin{figure*}[t]
	\centering
	\includegraphics[width=0.32\linewidth]{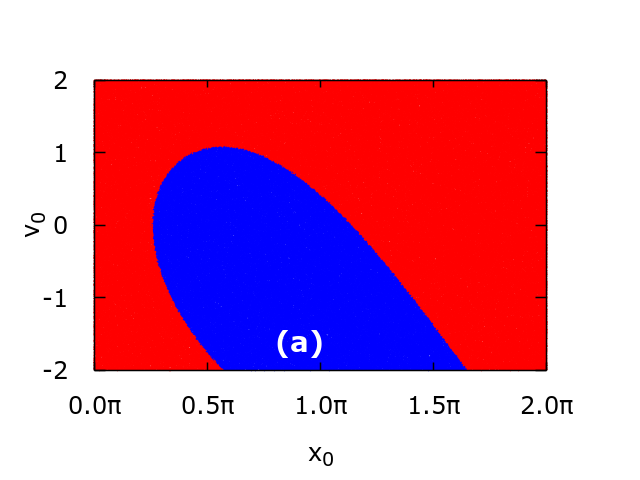}
	\includegraphics[width=0.32\linewidth]{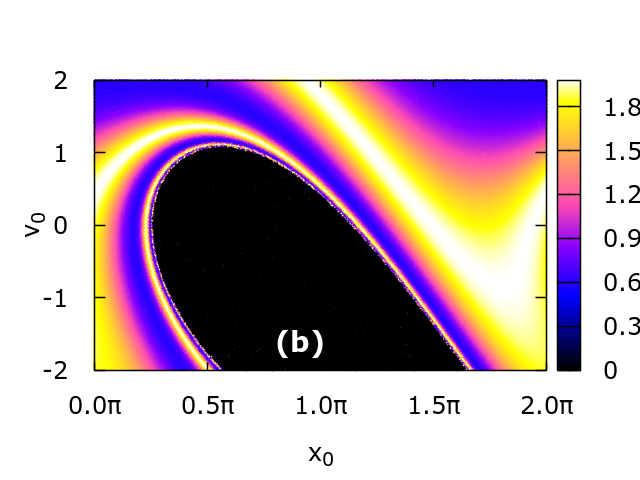}
	\includegraphics[width=0.32\linewidth]{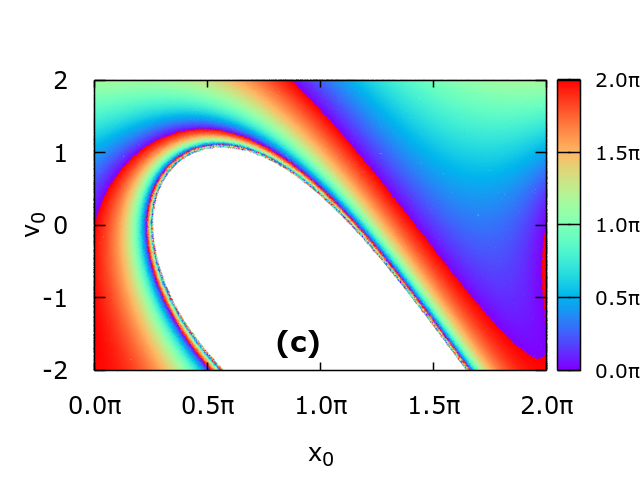}
	\caption{Panel (a): The fraction of basins of attraction for the time averaged velocity $\mathbf{v}$ of the particle. The blue region indicates the locked state whereas red stands for the running states. (b): Snapshot of the instantaneous velocity $v$ in the long time regime. (c): Phase difference $\phi$ between the trajectories $v(t)$ in the long time limit. All presented versus the initial conditions $\{x_0,v_0\}$ of the system. Other parameters read $\gamma = 0.66$, $f = 0.91$ and $\theta = 0$.}
	\label{fig1A}
\end{figure*}
\appendix
\section{Scaling of the  Langevin equation}
In the manuscript we investigate dynamics of a classical Brownian particle of mass $M$ dwelling in a spatially periodic and symmetric potential $U(x) = U(x + L)$ of the period $L$ and exposed to a constant force $F$. Such a system is described by the following Langevin equation
\begin{equation}
	\label{model}
	M\ddot{x} + \Gamma\dot{x} = -U'(x) + F + \sqrt{2\Gamma k_B T}\,\xi(t), 
\end{equation}
where the dot and prime denote differentiation with respect to the time $t$ and the particle coordinate $x$, respectively. The parameter $\Gamma$ is the friction coefficient and $k_B$ is the Boltzmann constant. The symmetric potential is assumed to be in the following simple form
\begin{equation}
	\label{potential}
	U(x) = -\Delta U \sin{\left(\frac{2\pi}{L}x\right)}
\end{equation}
where $\Delta U$ is half of the potential barrier height and $L$ is its spatial period. The coupling of the particle with thermal bath of temperature $T$ is modeled by the $\delta$-correlated Gaussian white noise $\xi(t)$ of vanishing mean and unit intensity, i.e.,
\begin{equation}
	\langle \xi(t) \rangle = 0, \quad \langle \xi(t)\xi(s) \rangle = \delta(t-s).
\end{equation}
The noise intensity factor $2\Gamma k_B T$ in Eq. (\ref{model}) follows from the fluctuation-dissipation theorem \cite{kubo1966} that ensures the canonical Gibbs statistics when the system is at the equilibrium state.

We define the dimensionless coordinate $\hat{x}$ and time $\hat{t}$ in the following way
\begin{equation}
	\label{scaling}
	\hat x = \frac{2\pi}{L} x, \quad \hat t = \frac{t}{\tau_0}, \quad 
	\tau_0 = \frac{L}{2\pi} \sqrt{\frac{M}{\Delta U}}, 
\end{equation}
where the characteristic time $\tau_0 = 1/\omega_0$ is the inverse of frequency $\omega_0$ of small oscillations in the potential well of $U(x)$. Under such a choice, Eq. (\ref{model}) is translated to the dimensionless form, namely, 
\begin{equation}
	\label{dimless-modelA}
	\ddot{\hat x} + \gamma \dot{\hat x} = -\mathcal{U}'(\hat x)+ \sqrt{2\gamma \theta}\,\hat{\xi}(\hat t),
\end{equation}
where now the dot and prime denote differentiation with respect to the dimensionless time $\hat t$ and coordinate $\hat x$, respectively. We note that the dimensionless mass is $m = 1$. The dimensionless friction coefficient $\gamma$ reads
\begin{equation}
	\gamma = \frac{\tau_0}{\tau_1} = \frac{1}{2\pi}\frac{L}{\sqrt{M \Delta U}}\, \Gamma,
\end{equation}
where the characteristic time $\tau_1 = M/\Gamma$ describes the velocity relaxation time for a free Brownian particle. 
The dimensionless total potential $\mathcal{U}(\hat x)$ takes the form 
\begin{equation}
\mathcal{U}({\hat x}) = -\sin {\hat x} -f {\hat x}. 
\end{equation}
The dimensionless force $f$ is given by
\begin{equation}
f = \frac{1}{2\pi}\frac{L}{\Delta U} F.
\end{equation}
The rescaled temperature $\theta$ is the ratio of thermal energy $k_{B} T$ to half of the activation energy the particle needs to overcome the original potential barrier $\Delta U$, i.e., 
\begin{equation}
	\theta = \frac{k_B T}{\Delta U}.
\end{equation}
The dimensionless thermal noise $\hat{\xi} (\hat t)$ is statistically equivalent to $\xi(t)$, namely, it is a Gaussian stochastic process with vanishing mean $\langle \hat{\xi}(\hat t) \rangle = 0$ and the correlation function \mbox{$\langle \hat{\xi}(\hat t) \hat{\xi}(\hat s) \rangle = \delta(\hat t-\hat s)$}. In the manuscript we use only the rescaled variables and therefore to improve the simplicity and readability of the notation we omit the hat appearing in Eq. (\ref{dimless-modelA}). 

\section{Ergodicity breaking}
One of the most important quantifier for characterizing the single particle transport is the time averaged velocity defined as
\begin{equation}
	\mathbf{v} = \lim_{t \to \infty} \frac{1}{t} \int_0^t ds \, \dot{x}(s).
\end{equation}
In the deterministic limit of vanishing thermal noise intensity $\theta \to 0$ ergodicity of the system given by \mbox{Eq. (\ref{dimless-model})} typically is strongly broken \cite{meroz2015,spiechowicz2016scirep}. It means that the phase space of the system decomposes onto mutually inaccessible and coexisting attractors for the mean particle velocity $\mathbf{v}$. Then different initial conditions $\{x_0,v_0\}$ for the particle coordinate and velocity lead to distinct averaged velocity $\mathbf{v}$ \cite{spiechowicz2017scirep,spiechowicz2019chaos}.
We exemplify this case in panel (a) of Fig. \ref{fig1A} where we depict a fraction of basins of attraction for  the time averaged velocity $\mathbf{v}$  when $\gamma = 0.66$ and $f = 0.91$. A larger set of the initial conditions is presented in Fig. 11.15 in Risken's book \cite{risken}. The initial phase space of the system is partitioned into two non-intersecting invariant regions corresponding to the locked and running states indicated by blue and red colors, respectively. Therefore ergodicity of the system is indeed broken. In the panel (b) we present the snapshot of the instantaneous long time velocity $v$ for the same parameter set to illustrate the complexity of the running solutions. All of them possess the same time average velocity (c.f. panel (a)) but in dependence on the initial conditions $\{x_0,v_0\}$ they assume different instantaneous values from the interval $[v_m,v_M]$, where $v_m = 0.59$ and $v_M = 1.98$. We note that for non-zero temperature $\theta$ the ergodicity of the system is restored and consequently the results are independent of any initial conditions \cite{kindermann2017}.

\section{Arcsine law}
For the paper to be self-contained we now recall the arcsine law for the random function
\begin{equation}  \label{VV}
	V(t) = A \sin{(\omega t + \phi)} + c, 
\end{equation}
where $A$ and $c$ are constants, and the phase $\phi$ is a random variable uniformly distributed on the interval $[0,2\pi]$, i.e. its probability density is  
\begin{equation} 	\label{uniform}
	g(\phi) = \frac{1}{2\pi} \,\theta(\phi) \,\theta(2\pi -\phi),  
\end{equation}
where $\theta(x)$ is the Heaviside step function. 
First, for the fixed time $t=t_i$ (time of sampling) we define the random variable $\eta = V(t_i)$.
Its characteristic function $\mathcal{C}_{\eta}(k)$ reads
\begin{eqnarray} 	\label{character}
	\mathcal{C}_{\eta}(k) &=& \langle \mbox{e}^{i k \eta} \rangle 
	= \int_{-\infty}^{\infty} \mbox{e}^{i k [A \sin(a+\phi) +c]} \; g(\phi) \; d\phi \nonumber \\ 
	&=&  \frac{1}{2\pi} \int_{0}^{2\pi} \mbox{e}^{i k [A \sin(a+ \phi) + c]} \; d\phi \nonumber \\ 
	&=& \frac{1}{2\pi} \int_{a}^{a+2\pi} \mbox{e}^{i k [A \sin(y) + c]} \; dy \nonumber \\ 
	&=&  \int_{-1}^{1} \mbox{e}^{i k (Az + c) } \frac{1}{\pi \sqrt{1-z^2}}\; dz \nonumber \\ 
	&=&  \int_{v_m}^{v_M} \mbox{e}^{i k u } \frac{1}{\pi \sqrt{(u-v_m)(v_M-u)}} \; du 	\nonumber \\ 
	&\equiv& \int_{-\infty}^{\infty} \mbox{e}^{i k u } \mathcal{P}_{\eta}(u) \; du.   
\end{eqnarray}
In the above equations we substituted $a = \omega t_i$ and introduced the minimal $v_m=c-A$ and maximal $v_M=c+A$ values of the function $V(t)$. 
From this relation it follows that the probability density of the random variable $\eta$ 
has the form
\begin{equation} 	\label{P}
	\mathcal{P}_{\eta}(u)  = \frac{\theta(u-v_m) \; \theta(v_M-u)}{\pi \sqrt{(u-v_m)(v_M-u)}}
\end{equation}
and the corresponding cumulative distribution defined on the interval $(v_m, v_M)$ 
is the arcsine function,  
\begin{equation} 	\label{cumu}
	\mathcal{F}(v) = \int_{v_m}^{v} {\mathcal P}_{\eta}(u) \; du = \frac{2}{\pi} \arcsin \sqrt{\frac{v-v_m}{v_M-v_m}}.
\end{equation}
For the system studied in the paper the values $v_M$ and $v_m$ are determined by two parameters, namely, the dimensionless friction coefficient $\gamma$ and the bias force $f$.

\end{document}